\documentclass[prl,floats,twocolumn]{revtex4}
\usepackage{graphicx}
\newcommand{\bk}{{\bf k}}
\newcommand{\bq}{{\bf q}}
\newcommand{\br}{{\bf r}}
\newcommand{\bR}{{\bf R}}
\newcommand{\bQ}{{\bf Q}}

\begin{document}
\title{Classical Antiferromagnetism in Kinetically Frustrated Electronic Models}
\author{C. N. Sposetti, B. Bravo, A. E. Trumper, C. J. Gazza, and L. O. Manuel}
\affiliation {Instituto de F\'{\i}sica Rosario (CONICET) and Universidad Nacional de Rosario,
Boulevard 27 de Febrero 210 bis, (2000) Rosario, Argentina} 
\vspace{4in }
\date{\today}

\begin{abstract} 
We study the infinite $U$ Hubbard model with one hole doped away half-filling, 
in triangular and square lattices with frustrated hoppings that invalidate Nagaoka's theorem, by means of the 
density matrix renormalization group. We find that these kinetically frustrated models have antiferromagnetic 
ground states with classical local magnetization in the thermodynamic limit. We identify the mechanism of this 
kinetic antiferromagnetism with the release of the kinetic energy frustration as the hole moves 
in the established antiferromagnetic background. This release can occurs in two different ways: 
by a non-trivial spin-Berry phase acquired by the hole or by the effective vanishing of the hopping amplitude along the
frustrating loops.
\end{abstract}
\maketitle

Itinerant magnetism has proved to be an elusive subject in condensed matter physics, since itinerant and 
localized aspects of electrons need to be taken into account on equal footing. 
The single-band Hubbard model, originally proposed to describe metallic 
ferromagnetism \cite{gutzwiller63}, has also been associated with antiferromagnetism of kinetic exchange 
origin close to half-filling. While {\it virtual} kinetic processes favor antiferromagnetism, it is a rule of thumb 
to link {\it real} kinetic processes with ferromagnetism \cite{fazekas99}. However, there exist only few exact results 
ensuring the existence of itinerant ferromagnetism \cite{nagaoka66,mielke91}. 
Among them, the most renowned is Nagaoka's theorem \cite{nagaoka66}, which assert that the 
saturated ferromagnetic state is the unique ground state when one hole is doped on the half-filled 
Hubbard model with infinite $U$ Coulomb repulsion. Furthermore, a connectivity condition must be fulfilled for the validity 
of Nagaoka's theorem: the sign of the hopping amplitudes around the smallest closed loop of the lattice 
must be positive, otherwise the hole kinetic energy will be frustrated and the saturated ferromagnetic state 
will no longer be the ground state. Kinetic energy frustration is a quantum mechanical phenomenon without classical analog, 
easily understood in certain tight-binding models where an electron can not gain 
the full kinetic energy $-z|t|,$ due to quantum interferences \cite{barford91,merino06}. This kind of 
frustration has been considerably less studied than the magnetic one, although recent works indicate 
that its effects may lead to rich physics, such as, robust superconductivity in strongly repulsive 
fermionic system \cite{isaev10} and spontaneous time-reversal symmetry breakings \cite{tieleman13}, 
among others \cite{wang08,yin11}. 

In a seminal work, Haerter and Shastry \cite{haerter05} 
have found a $120^{\circ}$ antiferromagnetic N\'eel order as the ground state of the $U=\infty$ 
triangular lattice Hubbard model when the hole motion is frustrated ($t > 0$), uncovering a new mechanism 
for itinerant magnetism. In this Letter, we further characterize this kinetic antiferromagnetism and we describe
its microscopic origin, analyzing generic kinetically frustrated electronic models for which, in 
the limit of infinite Coulomb repulsion and one hole doped away half-filling, 
the Nagaoka's theorem is not valid. In particular, we study the ground state of two Hubbard models: one on the 
triangular lattice with a positive hopping term, and the other on the 
square lattice with positive second-neighbor hopping term. Using the density matrix renormalization group
(DMRG) \cite{white01,white07}, we find in both cases that the ground state has antiferromagnetic 
order: $120^{\circ}$ N\'eel order for the triangular lattice and the usual $(\pi,\pi)$ 
N\'eel order for the square lattice. Surprisingly, we find that the local staggered 
magnetization becomes classical (saturated) in the thermodynamic limit. 
This result can be thought as the almost-perfect antiferromagnetic counterpart of the Nagaoka ferromagnetism; 
the difference is that, as the local staggered magnetization does not commute with the $SU(2)$ invariant Hubbard 
Hamiltonian, classical antiferromagnetic states can not be the exact eigenvectors for finite lattices. 
Based on a simple slave-fermion mean field \cite{manuel00}, we propose a mechanism responsible for 
the kinetic antiferromagnetism: if the hole were moving on a ferromagnetic background on these 
lattices, its kinetic energy would be frustrated. However, when moving in certain antiferromagnetic 
background, the hole can release its kinetic energy frustration by, depending on the system, 
acquiring a non-trivial spin Berry phase or having zero hopping amplitude along frustrating loops.
As the Coulomb repulsion is infinite, no exchange interaction exist, being the stabilization
of antiferromagnetism of pure kinetic origin.
 
{\it Hubbard model and DMRG.} ---We study the Hubbard model, 
$H=- \sum_{\langle ij\rangle \sigma}t_{ij}\left(\hat{c}^{\dagger}_{i\sigma}\hat{c}_{j \sigma} +\;{\rm H. c.}\right)+
U\sum_{i}\hat{n}_{i\uparrow}
\hat{n}_{i \downarrow},$
where we use the usual notation, and $\langle ij \rangle$ denotes pairs of neighbor sites connected by the hopping 
parameters $t_{ij}$. 
From the outset, we take $U=\infty.$  We study the Hubbard model on two lattices with frustrated kinetic hole energy: the triangular lattice 
with positive $t$ and the square lattice with nearest $t_1$- and positive next-nearest neighbor $t_2$ 
hopping terms. In the latter case, we choose $t_2=t_1 > 0$ as a generic point with kinetic frustration.  We take $t=t_1=1$ as the energy unit.

To solve the Hubbard model we apply DMRG on ladders of dimension $L_x\times L_y$ (see Fig. 1 
in \cite{weichselbaum11}), with up to $L_y=6$ legs and $L_x=15$ rungs. 
We choose clusters that are compatible with the antiferromagnetic orders found in this work.
This means that we take an even number of legs, $L_y$, for both lattices, and an even (multiple of 3) $L_x$ for 
the square (triangular) lattice. 
We consider cylindrical boundary conditions with periodic wrapping in the rung direction, and open boundary conditions 
along the legs. Comparing results obtained with clusters of different number of legs, we find that the clusters with 
$L_y=6$ give a correct description of the two-dimensional systems. A similar conclusion has been reached 
previously on square ladders \cite{liu12}. 
We have kept the truncation error less than $O(10^{-7})$, assuring that errors of the DMRG are smaller than symbol sizes in each figure.

We have also consider the inclusion of a weak pinning magnetic field ($B=0.1t$) 
acting only on a single site, at the lower left-hand end of the clusters \cite{liu12}, or on one magnetic sublattice. 
In previous works  \cite{white07}, the purpose of the 
small magnetic field has been to pin possible magnetic order, in order to reduce the computational efforts by computing the 
average value of the local spin instead of correlation functions, optimizing the truncation error. 
In our work, the inclusion of the weak magnetic field will allow us to highlight the classical character of the ground states.

\begin{figure}[ht]
\begin{center}
\includegraphics*[width=0.46\textwidth,angle=0]{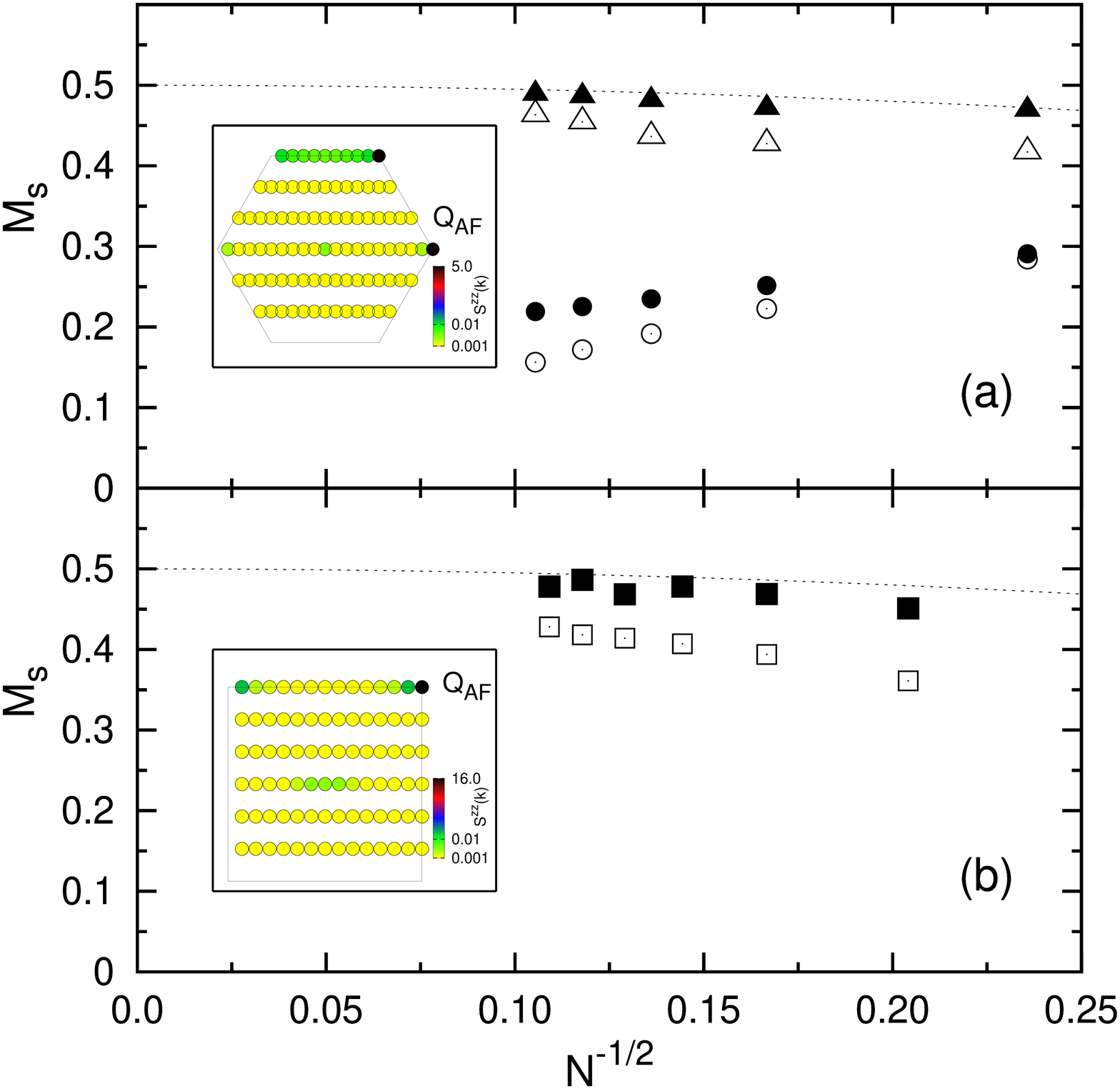}
\caption{(color online) Local magnetization versus $1/\sqrt{N}$ for (a) $U=\infty$ triangular Hubbard (triangles) and Heisenberg (circles) 
models, without magnetic field (open symbols) and with a magnetic field $B=0.1t$ applied to one sublattice (solid symbols); 
(b) $U=\infty$ square Hubbard model with first- and second-neighbor hopping terms ($t_1=t_2=1$), 
without magnetic field (open squares) and with a magnetic field $B=0.1t_1$ applied to one sublattice (solid squares). 
Dashed lines: classical local magnetization.
Insets: Intensity plot $S^{zz}(\bk)$ for the (a) triangular and (b) square models. Darker color indicates larger 
magnetic structure factor. $\bQ_{\rm AF}=(\frac{4\pi}{3},0)$ and $(\pi,\pi)$ for the triangular and square lattice, respectively.}
\label{fig1}
\end{center}
\end{figure}

{\it Static Magnetic Structure Factor.} ---To detect the existence of magnetic order, we compute the static magnetic 
structure factor with DMRG for both lattices, 
$S^{zz}(\bk)= \frac{1}{N}\sum_{ij}  \langle S^z_i\cdot S^z_j \rangle e^{-i\bk (\br_i - \br_j)},$ where $N=L_x L_y$ is the number of sites in the 
cluster, and $i,j$ run over all sites. 
In the inset of Fig. \ref{fig1}(a) we show an intensity plot of $S^{zz}(\bk),$ for a triangular cluster with $L_y=6$ legs and 
$N=90$ sites. For positive $t,$ $S^{zz}(\bk)$ exhibits two sharp maxima at the momenta ${\bQ}=\left(\frac{4\pi}{3},0\right)$ and 
${\bf Q}^*=\left(\frac{2\pi}{3},\frac{2\pi}{\sqrt{3}}\right),$ corresponding to a three-sublattice $120^{\circ}$ N\'eel order. 
As these peaks diverge with increasing cluster size, the ground state exhibits a long-range order $120^{\circ}$ 
magnetic pattern. This result
has been obtained previously by Haerter and Shastry\cite{haerter05}, diagonalizing an effective spin Hamiltonian on smaller clusters, 
up to $27$ sites.
Note that as we are working in the extremely correlated limit, where the exchange interaction driven by {\it virtual} kinetic 
processes vanishes, $J=0$, the magnetic order can only have its origin in the hole motion. On the other hand, as the triangular 
Heisenberg model has the same $120^{\circ}$ N\'eel order in its ground state \cite{bernu94,capriotti99}, for finite $U$ ($J > 0$) and low
doping, there is a synergy between {\it real} and {\it virtual} kinetic processes, that leads to the strengthening 
of the $120^{\circ}$ N\'eel order respect to the half-filled case \cite{weber06,sposetti14}. 

For negative $t$ (not shown in the figure) the magnetic structure factor has a sharp peak at $\bk=0$, while it vanishes for all 
other momenta. This result correspond to a fully polarized ferromagnetic ground state, as predicted by the Nagaoka's theorem. 

In the inset of Fig. \ref{fig1}(b) we show the magnetic structure factor for the $U=\infty$ Hubbard model on square clusters with
$L_y=6$ legs and $N=84$ sites, for $t_2=t_1 > 0$. 
Here, there is a marked peak for the magnetic wave vector ${\bQ}=\left(\pi,\pi\right)$, corresponding to the usual 
two-sublattice N\'eel order. 
If we relax the infinite $U$ condition, the kinetic exchange interactions would 
favor a collinear antiferromagnetic order, characterized by the magnetic wave vector $(\pi,0)$ or $(0,\pi)$, so there 
will be a competition between {\it real} and {\it virtual} kinetic processes, resulting in magnetic incommensuration and phase 
separation \cite{sposetti14}.  In agreement with Nagaoka's theorem, for  $t_2 < 0$ the ground state is the saturated ferromagnet.
 
{\it Local magnetization.} ---Once computed the magnetic structure factor, we can get the order parameter for 
the antiferromagnetic order, the local staggered magnetization 
$M_s=\sqrt{\frac{1}{N}\sum_{\alpha}\langle\left(\sum_{i \in \alpha}{S}_{i}\right)^2\rangle}$ \cite{nota2}, 
where $\alpha$ denotes the magnetic sublattices.  
The cluster-size dependence of $M_s$ for the $U=\infty$ triangular Hubbard model 
is shown in Fig. \ref{fig1}(a) (open triangles), along with the local magnetization of the 
triangular Heisenberg model (open circles) for comparison. 
The dashed line indicates the classical local 
magnetization, $M_{s,classic}=\frac{1}{2}-\frac{1}{2N},$ corrected by the presence
of the hole uniformly distributed (as is confirmed by the DMRG calculations).
Surprisingly, $M_s$ is very close to the classical value, even for small clusters, and it reaches this 
value in the thermodynamic limit ($L_y=6, L_x \to \infty$). 
On the other hand, in the Heisenberg case, strong zero-point quantum fluctuations 
lead to a drastic reduction of $M_s$ \cite{capriotti99}, and due to the quasi-one dimensional 
character of the clusters, $M_s$ extrapolates to 0 \cite{white07,weichselbaum11}. 
One possible reason for the classical character of the magnetic order is that the effective 
spin model, obtained after integrating the hole degree of freedom, contains effective long-range 
interactions (see Eq. 3 in Ref \cite{haerter05}) that may favor the classical ordering, like 
in the Lieb-Mattis model \cite{lieb62}. 

As we can see in Fig. \ref{fig1}(a), $M_s$ for finite size clusters does not 
take {\it exactly} the classical value. The Hubbard Hamiltonian is $SU(2)$ spin rotational 
invariant and it does not commute with the antiferromagnetic order parameter; consequently, 
its ground state can not break this symmetry for finite systems. Instead, it is expected that 
the finite size ground state is a singlet, and that, only in the thermodynamic limit, there 
can be a spontaneous $SU(2)$ symmetry breaking driven by the collapse of many low lying states 
onto the ground state \cite{bernu94}. The existence of a tower of states for the $U=\infty$ triangular 
Hubbard model has been confirmed in Ref. \cite{haerter05}. We argue that the small 
departure of the order parameter from the classical value is related with the singlet character of 
the finite size ground state, and not with zero-point quantum fluctuations that reduce the order parameter 
like in quantum antiferromagnets. Pictorially, the finite size ground state can be thought as a 
linear combination of several classical antiferromagnetic states lying in different planes.  
To strengthen this picture, we apply a small uniform pinning magnetic field, $B=0.1t,$ in one 
sublattice only (if we apply the magnetic field in only one site, the difference is 
quantitatively small, of only a few percent). 
Fig. \ref{fig1}(a) shows $M_s$ for the Hubbard model with the magnetic field applied (solid triangles) and 
the same for the triangular Heisenberg model (solid circles). It can be seen that in the Hubbard 
model now $M_s$ becomes classical, because the magnetic field select one of the classical orders that compose
the finite-size ground state. On the other hand, the magnetic field increases $M_s$ of the 
Heisenberg model, but there remains strong zero-point quantum fluctuations. 

Fig. \ref{fig1}(b) shows $M_s$ for the Hubbard model on the square lattice, 
with (solid squares) and without (open squares) an applied uniform magnetic field in one sublattice. The same
behavior as in the triangular case is found: the local magnetization is close to the classical values when $B=0$, being enough
to apply a small $B=0.1t$ to pin one classical magnetic ground state. 

{\it Energy scale.}---For the triangular lattice, the extrapolated ground state energy is $-4.178 \pm 0.001$, 
in agreement with the value obtained in \cite{haerter05} ($-4.183 \pm 0.005$), while for the square lattice the extrapolated 
value is $-4.848 \pm 0.001$. 
In order to quantify the energy scale of the kinetic antiferromagnetism, 
we match the effect of the hole motion to an effective nearest-neighbor antiferromagnetic Heisenberg interaction, 
$J_{eff}$\cite{haerter05,haerter06}, resulting $J_{eff} \simeq \Delta e\equiv (E_F-E_{AF})/N$, this is
the energy difference {\it per site} between the fully-polarized ferromagnetic state and the 
antiferromagnetic ground states. We have found that 
$J_{eff} \simeq  1.15/N$ ($J_{eff} \simeq 0.7/N$) for the triangular (square) lattice, for large 
$N.$ 

{\it Release of the kinetic frustration.} ---Now we trace 
back the origin of the kinetic antiferromagnetism by means of a comprehensive mean field approximation. 
To this end, we use the slave-fermion Schwinger-boson representation of the projected electronic 
degree of freedom  in the $t-J$ model, the strong-coupling limit of the Hubbard model.  
Here, we give a brief description of the mean field approach (see Supplemental Material for the details \cite{supplemental}).
In this representation, the projected electronic operator is written as 
$\tilde{c}_{i\sigma}=b^{\dagger}_{i\sigma}f_i,$ a composition of a Schwinger boson $b_{i\sigma}$, 
that account for the spin degrees of freedom, and a spinless slave fermion $f_i$, that describes 
the charge sector. 
This representation is replaced in the $t-J$ Hamiltonian, resulting
$H_{t\!-\!J} =  -\sum_{\langle ij\rangle} 2 t_{ij}\left({{\hat F}}_{ij} {{\hat B}}^{\dagger}_{ij} + h.c.\right) +
H_{bos}$  where we have defined the $SU(2)$ invariant operator 
${{\hat B}}^{\dagger}_{ij} = \frac{1}{2} \sum_\sigma b^{\dagger}_{i\sigma} b_{j\sigma},$ related with ferromagnetic 
correlations between sites $i$ and $j$ \cite{ceccatto93}, while
$\hat{F}_{ij}=f^{\dagger}_{i}f_j$ describes the hole hopping amplitude. 
$H_{bos}$ is a bosonic term that represents the spin fluctuations due to the Heisenberg term, and it vanishes when 
$J\to 0$ \cite{manuel00}. 
After a mean field decoupling, we get 
${H}_{t\!-\!J}^{MF} = \sum_{\bf k}\varepsilon_{f {\bf k}}{f}^{\dagger}_{\bf k} {f}_{\bf k} + H^{MF}_{bos},$  
where the hole kinetic energy dispersion takes the form $\varepsilon_{f {\bf k}} = 2\sum_{\bf R}t_R B_R \cos {\bk\cdot \bR}$
($\bf R$ are the relative position vectors of the sites connected by $t_R$). 

In the one hole case, the ground state energy of the system corresponds to the bottom of $\varepsilon_{f\bk}$. 
This energy dispersion is tight-binding like, with the hopping terms $t_R$ renormalized 
by the ferromagnetic mean-field parameter $B_R:$ $t_{R} \to t^{\rm eff}_R=t_R B_R$. 
The presence of the $B$'s parameters has two consequences: 
{\it i}) on one hand, the hopping terms are renormalized as in the double-exchange mechanism \cite{zener51},
$t^{\rm eff}_R \sim t_R \cos\frac{\varphi_R}{2}$ where $\varphi_R$ is the angle between the spins separated by vector $R$;  
if the spins are antiparallel $t^{\rm eff}_R$ vanishes; 
{\it ii}) on the other hand, the renormalization can give rise to a non-trivial spin Berry phase for non-collinear
orders, encoded in the $B_R$'s signs and associated with the solid angle subtended by the spins on a closed loop \cite{supplemental}. 

When the system is kinetically frustrated, these two features of the hopping renormalizations 
act releasing the hole kinetic energy frustration as the hole moves through certain antiferromagnetic patterns. 
Now we describe how this release works in the triangular and square lattice cases. {\it Triangular lattice:} 
for the ferromagnetic state all $B's$ parameters are equal to $S=1/2$, 
consequently the hole motion is frustrated for $t>0$. On the other hand, in a $120^{\circ}$ N\'eel order 
the parameters $B_{(\pm 1,0)}$ become negative and the others remain positive since
$B_R \sim M_s \cos \frac{\bQ \bR}{2}$ \cite{ceccatto93}. 
These negative $B$'s turn upside down the hole dispersion, releasing the kinetic frustration of the hole motion.
Notice that the flux of the $B$'s parameters in a closed loop is the solid angle 
subtended by the magnetic order, and consequently it is associated with the spin-Berry phase detected by the hole
(see Supplemental Material for details \cite{supplemental}). 
{\it Square lattice:} in this case, when $t_2 > 0$ the hole motion 
in a ferromagnetic state is frustrated. However, in the $(\pi,\pi)$ N\'eel order the vanishing of the effective 
first-neighbor hopping terms, due to their antiparallel spins \cite{zener51}, removes the frustrating loops, releasing the
kinetic frustration. 

We remark that, at the mean field level, the 120$^{\circ}$ N\'eel ($(\pi,\pi)$ N\'eel) state is degenerate with 
the  ferromagnetic one in the triangular case with $t>0$ (square lattice with $t_2>0$). 
The reason for this degeneracy is that, although the hole motion through the antiferromagnetic states is not 
frustrated, there is a hole dispersion bandwidth reduction due to the hopping renormalizations \cite{supplemental}. 
So, strictly speaking, the mean field numerics does not show the stabilization of the kinetic antiferromagnetism over
the Nagaoka's state. However, after considering the combined effects of quantum interference and strong-correlation 
physics beyond the mean field approximation (like our DMRG predictions), the actual kinetic antiferromagnetism emerges. 
Despite the mean field discrepancy, we strongly emphasize that the mean field approach allows to find one of the main 
ingredient of the kinetic antiferromagnetism, that is, the release of the kinetic frustration. 

Using the insight we gained from the mean field approach, we can predict the appearence of 
this novel kinetic antiferromagnetism phenomenon in other kinetically frustrated systems, 
like the anisotropic triangular lattice and the $t_1-t_2$ square lattice with 
$t_2 \neq |t_1|.$ Preliminary DMRG results show the ubiquity of kinetic antiferromagnetism in these systems (see Supplemmental 
Material \cite{supplemental}). Furthermore, we remark that the kinetic antiferromagnetism 
mechanism is completely different to the exchange one, and in general the ground state selected 
by this itinerant mechanism does not have to be necessarily the classical ground state of the related Heisenberg model. 
In this context, in particular, the $U=\infty$ kagom\'e Hubbard model with one hole doped and $t>0,$ 
may be a promising candidate for the search of unconventional kinetic antiferromagnetism physics \cite{supplemental}. 

\begin{figure}[ht]
\begin{center}
\includegraphics*[width=0.46\textwidth]{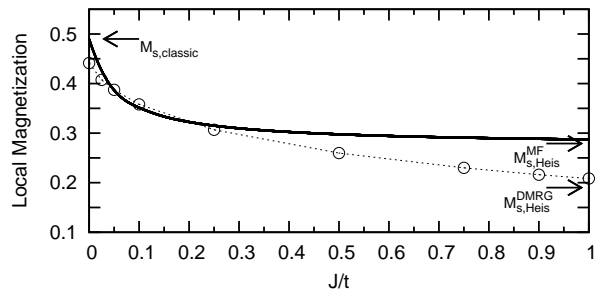}
\caption{Local magnetization $M_s$ of the triangular $t-J$ model as a function of $J/t$ for doping $\delta=0.0185.$ 
The solid line corresponds to the mean field $M_s$, the open circles to the DMRG results, the dashed line is to aid the eye.
The arrows indicate the $J=0$ and Heisenberg $M_s$ limits.}
\label{fig2}
\end{center}
\end{figure}

Finally, if we lift the condition of infinite $U,$ it is possible to study the 
synergy between {\it real} and {\it virtual} kinetic processes, in order to highlight 
the crossover from the Heisenberg regime, governed by the 
exchange interactions, to the kinetic antiferromagnetic one, governed by the kinetic energy. 
In Fig. \ref{fig2}  we show the local magnetization of the ground state $120^{\circ}$ N\'eel order of 
the triangular $t-J$ model predicted by the mean field approach and DMRG \cite{note3}, as a function of $J/t$ for doping $\delta=0.0185$. 
There is fairly good qualitative agreement between both methods, 
and for larger values of $J/t$ the order parameter is close to the Heisenberg value calculated within each approach, 
$M^{\rm MF}_{s,Heis} \sim 0.275$ \cite{gazza93} and $M^{\rm DMRG}_{s,Heis} \sim 0.205$ \cite{white07}, while $M_s$ increases with decreasing $J/t$, 
until it reaches the classical value for $J/t \to 0$ in both methods.

{\it Conclusions.} ---Using the density matrix renormalization group, 
we find that classical antiferromagnetic ground states can be realized in extremely correlated
electronic systems with frustrated kinetic energy. In particular, we study the $U=\infty$ 
Hubbard model, with one hole doped
away half filling, on the positive $t$ triangular lattice, and  on the square lattice with 
positive second neighbor hoppings. 
We also propose a mechanism responsible for this kinetic antiferromagnetism, that is, 
the release of the kinetic energy frustration driven by, depending on the system, the spin-Berry 
phase acquired by the hole while moving around an antiferromagnetic background, or
the vanishing of the effective hopping amplitude along the frustrating loops.
This new mechanism for itinerant antiferromagnetism is quite ubiquitous 
for one hole doped away half filling in kinetically frustrated lattices \cite{supplemental}, 
being also relevant in more general situations, like finite $U$ regime and 
low doping cases, as the mean-field results seem to indicate \cite{sposetti14}. 
It is worth noticing that recent experiments \cite{struck11} 
were able to generate gauge fields which induce frustrated motion of ultracold bosons in 
triangular optical lattices, opening up the possibility to observe related kinetic 
antiferromagnetism phenomena.

We acknowledge useful discussions with C. D. Batista and D. J. Garc\'ia. This work was partially 
supported by PIP CONICET Grants 0160 and 0392.

\newpage
\widetext

\section{Supplementary material}

Here we present the slave-fermion mean-field results for the $t-J$ model with $J=0$, that allow us to interpret the physical origin of the kinetic
antiferromagnetism found numerically with the DMRG method.

\section{Slave fermion mean field approximation}
First, we briefly introduce the slave-fermion mean-field treatment  \cite{jayaprakash89} of the $t-J$ model,
\begin{equation}
 \hat{H}_{tJ}=-\sum_{i R \sigma}t_{R}\hat{\tilde{c}}^{\dagger}_{i\sigma}\hat{\tilde{c}}_{i+R\sigma}+
 \frac{1}{2}\sum_{i R}J_{R}{\bf S}_i\cdot {\bf S}_{i+R},
 \label{tJ}
\end{equation}
where the sum $\sum_{i R}$ is over all the $i$ sites of the cluster and all the $R$ neighbors connected with $i$ 
by the exchange interaction $J_{R}$ and hopping term $t_{R}.$ The $t-J$ model is the strong coupling 
$U/t_R \gg 1$ limit of the Hubbard model, with $J_R=4t_R^2/U$, and it is written in terms of the projected electronic operators,  
$\tilde{\hat{c}}_{i\sigma} \equiv \hat{c}_{i\sigma}\left(1-\hat{n}_{i -\sigma}\right),$ with $\hat{n}_{i-\sigma} 
= \hat{c}^{\dagger}_{i-\sigma}\hat{c}_{i -\sigma},$ forbidding the electronic double occupancy at each site.
For $U=\infty$, $J_R=0$, and both models are identical.

The projected electronic operators can be represented in term of Schwinger bosons $\hat{b}_{i\sigma}$ and spinless 
fermions $\hat{f}_i:$ \begin{equation}
 \hat{\tilde{c}}_{i\sigma}=\hat{f}^{\dagger}_i \hat{b}_{i\sigma}.
 \label{cproj}
\end{equation}
This replacement, together with the constraint 
\begin{equation}
\sum_{\sigma}\hat{b}^{\dagger}_{i\sigma}\hat{b}_{i\sigma}+\hat{f}^{\dagger}_{i}\hat{f}_i=1
\label{const}
\end{equation}
at each site $i$, is a faithful representation of the original Fermi algebra. The spin can be written in terms of the Schwinger bosons as
${\bf S}_i=\sum_{\sigma \sigma'}\hat{b}^{\dagger}_{i\sigma}{\vec \sigma}_{\sigma \sigma'}\hat{b}_{i\sigma'}$, where ${\vec \sigma}$ is the vector of Pauli matrices, 
while the slave fermion $\hat{f}_i$ corresponds to the charge (hole) degree of freedom.

We replace (\ref{cproj}) in the $t-J$ Hamiltonian (\ref{tJ}) and we obtain
\begin{equation}
 \hat{H}_{tJ}=\sum_{i R}t_{R} \hat{F}_{i R}\hat{B}^{\dagger}_{i R}+\frac{1}{2} \sum_{i R}J_{R}
 \left[:\hat{B}^{\dagger}_{i R}\hat{B}_{i R}:-\hat{A}^{\dagger}_{i R}\hat{A}_{i R}\right],
 \label{tjslave}
\end{equation}
where we have defined the singlet $SU(2)$ operators
${\hat A}^{\dagger}_{i R}=\frac{1}{2}\sum_{\sigma}\sigma \hat{b}^{\dagger}_{i\sigma}\hat{b}^{\dagger}_{i+R -\sigma}$ and 
${\hat B}^{\dagger}_{iR}=\frac{1}{2}\sum_{\sigma}\hat{b}^{\dagger}_{i\sigma}\hat{b}_{i+R \sigma}$ \cite{ceccatto93}, while 
$\hat{F}_{iR}=\hat{f}^{\dagger}_{i+R}\hat{f}_i$. 
${\hat A}_{iR}$ (${\hat B}_{iR}$) is related with the antiferromagnetic (ferromagnetic) correlations between sites $i$ and $i+R$, while ${\hat F}_{iR}$ represents 
the probability that a hole hopes from $i$ to $i+R$. 

Hamiltonian (\ref{tjslave}) involves quartic terms in slave operators, and thus we appeal to a mean-field decoupling \cite{ceccatto93} in order to solve it approximately. 
Also, we approximate the local constraint (\ref{const}) by its average over all the lattice, enforced by a Lagrange multiplier $\lambda$. 
The resulting mean-field Hamiltonian is 
\begin{equation}
\hat{H}^{MF}_{tJ}=\sum_{iR}t_R B^*_R \hat{F}_{iR}+\sum_{iR}t_R F_R\hat{B}^{\dagger}_{iR}+\sum_{iR}\frac{J_R}{2}\left(B_R \hat{B}^{\dagger}_{iR}- 
A_R \hat{A}^{\dagger}_{iR}+{\rm H.c.}\right)+\lambda \sum_{i\sigma}\hat{b}^{\dagger}_{i\sigma}\hat{b}_{i \sigma}+\mu \sum_i \hat{f}^{\dagger}_i \hat{f}_i+{\rm Cte.}
\end{equation}
To obtain this expression we replace the bilinear operators $\hat{A}$, $\hat{B}$ and $\hat{F}$ by its (real) averages, assuming a translational invariant solution. Also, 
we introduce the chemical potential $\mu$ to control the hole doping $\delta$.
After transforming to the Fourier space, we get
\begin{equation}
 \hat{H}^{MF}_{tJ}=\sum_{\bk}\varepsilon_{f\bk}\hat{f}^{\dagger}_{\bk}\hat{f}_{\bk}+\sum_{\bk}
 \left[\varepsilon_{b\bk}\left(\hat{b}^{\dagger}_{\bk\uparrow}\hat{b}_{\bk\uparrow}+
 \hat{b}^{\dagger}_{-\bk \downarrow}\hat{b}_{-\bk \downarrow}\right)+i\gamma_{\bk}\left(\hat{b}^{\dagger}_{\bk \uparrow}\hat{b}^{\dagger}_{-\bk \downarrow}-
 \hat{b}_{-\bk \downarrow}\hat{b}_{\bk \uparrow}\right) \right] + {\rm Cte.},
 \label{tjk}
\end{equation}
where the hole dispersion is 
\begin{equation}
 \varepsilon_{f\bk}=2\sum_R t_R B_R \cos \bk \bR + \mu,
 \label{edisp}
\end{equation}
while $\gamma_{\bk}=\frac{1}{2}\sum_R J_R A_R \sin \bk \bR$ and $\varepsilon_{b\bk}=\sum_{R}\left(t_R F_R + \frac{1}{2}J_R B_R\right)\cos\bk \bR + \lambda.$ 

Now we have a quadratic Hamiltonian (\ref{tjk}), that can be straightforwardly diagonalized by means of a Bogoliubov transformation of the Schwinger bosons to new bosonic 
operators $\hat{\alpha}_{\bk \sigma}$'s \cite{mezio13}:
\begin{equation}
 \hat{H}^{MF}_{tJ}=\sum_{\bk}\varepsilon_{f\bk}\hat{f}^{\dagger}_{\bk}\hat{f}_{\bk}+\sum_{\bk \sigma}\omega_{b\bk}\hat{\alpha}^{\dagger}_{\bk \sigma}\hat{\alpha}_{\bk
 \sigma} + {\rm Cte.},
 \label{mf}
\end{equation}
where the boson dispersion is $\omega_{\bk \sigma}=\sqrt{\varepsilon_{b\bk}^2-\gamma_{\bk}^2}.$ Finally, 
the mean field parameters $A_R$, $B_R,$ and $F_R$ can be computed in a self consistent way \cite{mezio13}.

The mean-field Hamiltonian (\ref{mf}) approximates the complicated $t-J$ model by two one-body terms: a fermionic tight-binding-like one, describing the motion of holes, 
and a bosonic term, corresponding to the spin fluctuations of the system. It should be noticed that both terms are interdependent, as the hole dispersion depends on 
the ferromagnetic parameter $B_R$ while the bosonic dispersion depends on $F_R$, providing a coupling between the charge and spin degrees of freedom.

Once we obtain the mean field Hamiltonian, it is a simple task to calculate the 
ground state physical properties of the system as a function of $t_R$, $J_R,$ and hole doping, $\delta.$ In particular, 
the existence of magnetic order is manifested in the Bose-Einstein condensation of the bosons at $\bk=\pm \bQ/2$, where $\bQ$ is the magnetic wave vector 
\cite{sarker89}. In Fig. 2 of the manuscript we show the mean field local magnetization of the triangular $t-J$ model as a function of $J/t,$ 
in order to illustrate the 
ordering effect of kinetic antiferromagnetism in this lattice. 
In a future publication \cite{sposetti14}, we will present a detailed study of the mean field phase diagram 
of the $t-J$ model on the triangular and square lattices. Hereafter, 
we report the mean field results for the special case of interest in this work, that is, $J=0$.

\section{Results for $J=0$}

\begin{figure}[t] 
\begin{center}
\includegraphics[width=7cm]{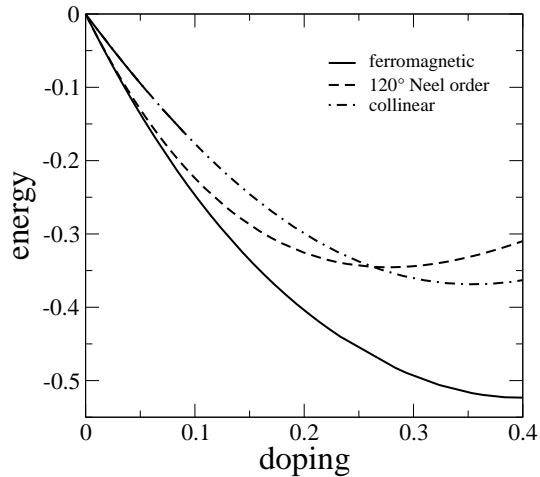}
\caption{Slave-fermion mean-field energy {\it per site} as a function of doping for $120^{\circ}$ N\'eel, ferromagnetic, and the collinear stripe order, for 
$J=0$ and $t > 0$ in the triangular lattice.}
\label{enertpos}
\end{center}
\end{figure}

For $J=0$, the $\gamma_{\bk}$ factors vanish identically, consequently Hamiltonian (\ref{tjk}) is already diagonal and 
there is no need of a Bogoliubov transformation. Instead, it is imperative to condense the Schwinger bosons, in order to satisfy the averaged constraint 
(\ref{const}). 
Considering Bose condensation at different momenta $\bk=\pm \frac{\bQ}{2}$, we can calculate the energy of magnetic phases (if the phases are locally stable, that is, 
they satisfy the self-consistent mean-field equations) 
characterized by different magnetic wave vectors $\bQ$'s. In particular, for the triangular lattice and $t>0$, we evaluate the energy of the 
ferromagnetic phase ($\bQ=(0,0)$), the $120^{\circ}$ N\'eel order ($\bQ=(\frac{4\pi}{3},0)$), and a collinear stripe order ($\bQ=(0,\frac{2\pi}{\sqrt{3}})$), 
that is favorable for intermediate doping and $J/t$ \cite{sposetti14,weber06}. 

In Fig. \ref{enertpos} we show the energy {\it per site} of the three phases as a function of hole doping, calculated for a large cluster ($N=30000$ sites) that 
represents the thermodynamic limit. It can be seen that, at the mean-field level, the ferromagnetic phase is the ground state for all the doping range considered, 
although the 120$^{\circ}$ N\'eel order
has a very close energy for doping $\delta \lesssim 0.05$. The collinear stripe phase always has a much larger energy, and so it is not stabilized by 
the kinetic energy alone. It should be stressed that after an exhaustive search for solutions of the mean-field equations for finite doping, 
we have always found that the ferromagnetic phase is the ground state.
In the case of only one hole doped away half-filling, we will show in the next section that the ferromagnetic phase is degenerate with the $120^{\circ}$ N\'eel order. 
The energy of both phases is $E^{MF}=-3t$; while this is the exact ferromagnetic phase energy, 
it is clearly rather above the exact 120$^{\circ}$ N\'eel phase energy found by our DMRG calculations and exact diagonalization \cite{haerter05}. It is worthy to note
that, once $J/t > 0$, the 120$^{\circ}$ N\'eel state is strongly favored over the ferromagnetic one \cite{sposetti14}, because the exchange interaction select 
the 120$^{\circ}$ magnetic pattern in the triangular lattice \cite{manuel98}.

\begin{figure}[t] 
\begin{center}
\includegraphics[width=7cm]{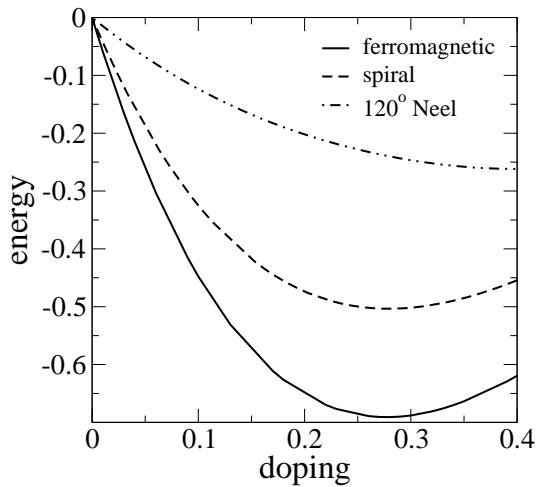}
\caption{ Slave-fermion mean-field energy {\it per site} as a function of doping for $120^{\circ}$ N\'eel, ferromagnetic, and spiral $\bQ=(\frac{2\pi}{3},0)$ phases, for 
$J=0$ and $t < 0$ in the triangular lattice.}
\label{enertneg}
\end{center}
\end{figure}

For $t<0$, it can be seen in Fig. \ref{enertneg} that the ferromagnetic phase is also the ground state for all doping, but now its energy 
is well separated from the energies of the other phases considered (120$^{\circ}$ N\'eel phase and a representative spiral phase, with magnetic wave vector 
$\bQ=(\frac{2\pi}{3},0)$). For negative $t$ Nagaoka's theorem is valid, and it is believed that ferromagnetism is very stable for further doping 
\cite{koretsune02}.

So, for $t>0$ and $J=0,$ the mean-field approximation does not capture that $120^{\circ}$ N\'eel order is the unique ground state, as it is degenerate with the 
ferromagnetic phase. This failure has to do with the mean-field character of the approach, in which we approximate the actual many-body 
problem by one-body terms. From previous experience  \cite{trumper97,manuel98,manuel99}, 
we expect that the inclusion of Gaussian fluctuations above the mean-field solutions 
will put the slave-fermion predictions in consonance with our DMRG and exact diagonalization \cite{haerter05} results, 
lowering the 120$^{\circ}$ N\'eel energy while leaving unchanged the already exact ferromagnetic energy. 

For the $t-J$ model on the square lattice with first and second neighbor hoppings, with $J_R=0$ and $|t_1|=|t_2|$, we obtained analogous results to the triangular lattice. 
For $t_2 < 0$, the ferromagnetic phase is clearly the ground state, in accordance with the Nagaoka's theorem. 
For $t_2 > 0,$ although the ferromagnetic phase is the ground state for all doping considered ($\delta < 0.40$), its energy is very close to the energy 
of the usual $\bQ=(\pi,\pi)$ N\'eel state, being both states degenerate for one hole doped away half-filling. Again, this result does not agree with the DMRG 
prediction of a unique antiferromagnetic ground state.

Although the mean-field theory fails to obtain the correct antiferromagnetic ground states for $J=0$, 
in the next section we argue that it is a very helpful approach to elucidate the physical mechanism behind the phenomenon of kinetic antiferromagnetism.

\section{Release of kinetic energy frustration}

In this section we limit ourselves to the case of one hole doped, relevant for our work. We analyze 
why the 120$^{\circ}$ N\'eel order ($\bQ=(\pi,\pi)$) is one of the ground states, along with the ferromagnetic one, of the mean field Hamiltonian (\ref{tjk}) for the triangular 
lattice (square lattice).
The answer lies in the release of the hole kinetic energy frustration thanks to the hole motion on the antiferromagnetic background, 
and we believe that this mechanism can be used to interpret the kinetic antiferromagnetism. 
In the one hole case, the ground state energy of the system corresponds to the energy of the bottom of the hole dispersion, 
$\varepsilon_{f\bk}$ (\ref{edisp}). This energy dispersion is a tight-binding like, with the hopping terms $t_R$ renormalized 
by the ferromagnetic mean-field parameter $B_R:$ $t_{R} \to t^{\rm eff}_R=t_R B_R$. On one hand, this renormalization can induce a 
spin-Berry phase, while, on the other hand, it could lead to the vanishing of the hopping amplitude between certain neighbor sites. 
Although, the mechanism is the same for both lattices, the details differ, so we analyze each case separately.

\subsection{Triangular lattice}
In the triangular lattice, the hole dispersion (\ref{edisp}) reads 
\begin{equation}
\varepsilon_{f {\bf k}} = 4 t \left[ B_1 \cos k_x + B_2 \cos \left(\frac{k_x}{2} + \frac{\sqrt{3}k_y}{2}\right) 
                            + B_3 \cos\left( -\frac{k_x}{2}+\frac{\sqrt{3}k_y}{2} \right) \right],
\label{rdh}
\end{equation}
where the parameters $B_i$'s correspond to the neighbors $R_1=(1,0)$, $R_2=(\frac{1}{2},\frac{\sqrt{3}}{2}),$ and $R_3=(-\frac{1}{2},\frac{\sqrt{3}}{2}).$

For $J=0,$ the analysis of the mean-field equations shows that the $B_R$ parameters take their 
classical values, $B_R=\frac{1}{2}\cos \frac{\bQ \bR}{2}$ \cite{ceccatto93,note1}.
So, for the ferromagnetic phase, $\bQ=(0,0)$, the hole dispersion takes the form
\begin{equation}
 \varepsilon_{f\bk}^{ferro}=2 t \left[\cos k_x + 2\cos \frac{k_x}{2} \cos \frac{\sqrt{3}k_y}{2}\right],
 \end{equation}
while for the 120$^{\circ}$ N\'eel order, $\bQ=\left(\frac{4\pi}{3},0\right)$,
\begin{equation}
 \varepsilon_{f\bk}^{120^{\circ}}=t \left[-\cos k_x + 2\cos \frac{k_x}{2} \cos \frac{\sqrt{3}k_y}{2}\right].
 \end{equation}

The minimum and maximum (maximum and minimum) of the ferromagnetic energy dispersion occur at the momenta 
${\bf k}=(\frac{2\pi}{3},\frac{2\pi}{\sqrt{3}})$ and $(0,0)$, respectively, for $t > 0$ ($t<0$):
\begin{eqnarray*}
 \varepsilon_{f,(\frac{2\pi}{3},\frac{2\pi}{\sqrt{3}})} &=& -3t \\
 \varepsilon_{f,(0,0)} &=&  6  t
\end{eqnarray*}
As a consequence, the hole motion is frustrated (unfrustrated) for $t>0$ ($t<0$).

For the $120^{\circ}$ N\'eel phase, the minimum and maximum (maximum and minimum) now occur at the momenta 
${\bf k}=$ $(0,\frac{2\pi}{\sqrt{3}})$ and $(\pm\frac{2\pi}{3},0)$, respectively, for $t>0$ ($t < 0$). 
\begin{eqnarray*}
 \varepsilon_{f,(0,\frac{2\pi}{\sqrt{3}})} &=& -3 t\\
 \varepsilon_{f,(\pm\frac{2\pi}{3},0)} &=& \frac{3t}{2}
\end{eqnarray*} 

In the 120$^{\circ}$ N\'eel phase, on one hand, the hole dispersion bandwidth is halved. This results from 
the fact that the effective hopping $t^{\rm eff}_R$ between sites $i$ and $i+R$ 
is proportional, through the modulus of $B_R$, to the overlap of the spin 
wavefunctions pointing in the direction of the local magnetization at each site, 
similar to what happens in the double exchange mechanism \cite{zener51}, $t^{\rm eff}_R \propto \cos\frac{\varphi_R}{2}$ where 
$\varphi_R$ is the angle between spins at sites $i$ and $i+R$.
On the other hand, the negative sign acquired by the $B_1$ parameter turns upside  
down the hole dispersion with respect to the ferromagnetic case. 

As the energy of $t-J$ model, with $J=0$ and one hole doped away half-filling, coincides with the bottom of the hole dispersion energy, 
for $t<0$ the ferromagnetic phase is the ground state, with a significantly lower energy than the 120$^{\circ}$ N\'eel state.
The mean-field approximation gets the exact ground state energy, $E^{\rm ferro}=-6|t|$, and it satisfies Nagaoka's theorem. 
As it is shown in Fig. (\ref{enertneg}) ferromagnetism clearly prevails for further doping.
In contrast, for $t>0,$ the hole motion in the ferromagnetic state is frustrated, 
while the kinetic energy frustration is released in the 120$^{\circ}$ state thanks 
to the negative $B_1$ parameter. 
At the mean-field level this release gives rise to the degeneracy of the ferromagnetic and the 120$^{\circ}$ N\'eel phases.
While the mean-field ferromagnetic energy is the exact one ($E^{\rm ferro}=-3t$), the mean-field energy 
of the 120$^{\circ}$ state is approximate, and it is necessary to use more sophisticated techniques beyond mean-field, 
like the DMRG computations we perform in our work or the exact diagonalization by Haerter and Shastry \cite{haerter05}, 
to capture the stabilization of the 120$^{\circ}$ phase over
the ferromagnetic one. It is worth to notice that the energy difference between the ferromagnetic and the 120$^{\circ}$ phases computed by DMRG, 
$\Delta E = 1.178 t$, is relatively large ($\sim 40 \%$ of the ferromagnetic energy), indicating that many-body effects beyond mean-field are
really important.

In spite of this discrepancy of the mean-field approximation, we believe that it is a very helpful approach 
because we can extract from it the main ingredient of the physical mechanism that gives rise to the kinetic antiferromagnetism 
in these extremely correlated systems: 
in kinetically frustrated situations --when Nagaoka's theorem is not valid--, the kinetic 
frustration can be released if the hole moves in certain antiferromagnetic backgrounds, 
lowering its energy with respect to the ferromagnetic state (when quantum fluctuations beyond mean-field are considered). 

The sum of the (complex) phases of the $B_R$ parameters around a closed loop is the spin-Berry phase acquired by the hole when 
it runs through the loop, and it is also half the solid angle subtended by the spins on the loop \cite{misguich05}. 
While in the ferromagnetic phase (like in any collinear phase) the subtended solid angle vanishes, in the 120$^{\circ}$ N\'eel phase 
this angle is $2\pi$ for each unit triangle, and its associated spin-Berry phase is $\pi$ (related to the negative sign of $B_1$  \cite{note2}). 
So, we can summarize that in the triangular lattice the mechanism of kinetic antiferromagnetism consists in the release of the kinetic frustration 
by means of the spin-Berry phase acquired by the hole as it moves along the 120$^{\circ}$ magnetic pattern. 

\subsection{Square lattice with $|t_1|=|t_2|$}

In the square lattice with hopping terms to first and second neighbors, the hole dispersion (\ref{edisp}) is 
\begin{equation}
 \varepsilon_{f\bk}=4t_1\left[B_1 \cos k_x + B_2 \cos k_y \right] + 4t_2\left[B_3 \cos\left(k_x+k_y\right)+
 B_4 \cos\left(-k_x+k_y\right)\right],
\end{equation}
where the $B_i$'s parameters correspond to the first neighbors $R_1=(1,0)$ and $R_2=(0,1)$, and to the second neighbors, $R_3=(1,1)$ and $R_4=(-1,1)$.

In the ferromagnetic phase, again $B_R=1/2$ for all $R$ \cite{note1}, and the hole dispersion becomes
\begin{equation}
 \varepsilon^{\rm ferro}_{f\bk}=2t_1\left[\cos k_x + \cos k_y \right] + 4t_2\cos k_x \cos k_y.
\end{equation}

As the $t_1$ sign is irrelevant, we take $t_1>0$ in the following. For $|t_2|= t_1$, 
the minimum (maximum) of the ferromagnetic energy dispersion occurs at the momenta, 
${\bf k}=(\pi,0)$ or $(0,\pi)$ ($\bk=(0,0)$) for $t_2 > 0:$
\begin{eqnarray*}
 \varepsilon_{f,(\pi,0) {\rm or}(0,\pi)} &=& -4t_2 ,\\
 \varepsilon_{f,(0,0)} &=&  4 t_1+ 4t_2; 
\end{eqnarray*}
while for $t_2 < 0$ the minimum (maximum) occurs at the momenta 
${\bf k}=(\pi,\pi)$ (${\bf k}=(\pi,0)$ or $(0,\pi)$):
\begin{eqnarray*}
 \varepsilon_{f,(\pi,\pi)} &=& -4t_1+4t_2, \\
 \varepsilon_{f,(\pi,0) {\rm or}(0,\pi)} &=& -4t_2 , 
\end{eqnarray*}

As a consequence, the hole motion in the ferromagnetic background is frustrated (unfrustrated) for $t_2>0$ ($t_2<0$).

For the $\bQ=(\pi,\pi)$ N\'eel phase, the first neighbor $B_R$ parameters vanish identically because of the antiparallel 
orientation of the spins, while $B_3=-\frac{1}{2}$ and $B_4=\frac{1}{2}$. Therefore, the hole dispersion is 
\begin{equation}
 \varepsilon^{\rm (\pi,\pi)}_{f\bk}=4 t_2\sin k_x \sin k_y.
\end{equation}
As it is well known for this kind of mean-field approximations, the holes can only propagate along the ferromagnetic sublattices.
The minimum and maximum (maximum and minimum) of the hole dispersion now are located at momenta $\bk=\pm\left(\frac{\pi}{2},-\frac{\pi}{2}\right)$ 
and $\bk=\pm\left(\frac{\pi}{2},\frac{\pi}{2}\right)$, respectively, for $t_2 > 0$ ($t_2 < 0$), with 
\begin{eqnarray*}
 \varepsilon_{f,\pm(\frac{\pi}{2},-\frac{\pi}{2})} &=& -4 t_2,\\
 \varepsilon_{f,\pm(\frac{\pi}{2},\frac{\pi}{2})} &=& 4 t_2.
\end{eqnarray*} 

For $t_2 < 0$, the unique ground state of the $t-J$ model, with $J=0$ and one-hole doped away half-filling, 
is the ferromagnetic phase, with the optimal kinetic energy $E^{\rm ferro}=-8|t_1|,$ well separated from the N\'eel phase energy ($E^{\rm Neel}_{MF}=-4|t_1|$). 
As in the triangular case, the mean-field approximation gives the exact ground state energy and satisfies Nagaoka's theorem. 

For $t_2 > 0$ the ferromagnetic and N\'eel states are degenerate at the mean field level ($E^{\rm ferro}=E^{\rm Neel}_{MF}=-4|t_1|$).  
As in the triangular case, the ferromagnetic phase mean-field energy is the exact one, while the mean-field N\'eel energy is approximate. 
Our DMRG computations shows that the N\'eel phase is the true ground state, with an energy $E^{\rm Neel}=-4.848 |t_1|,$ well below the mean-field value. 

In the square lattice model, we can also conclude that the origin of the kinetic antiferromagnetism is intimately related with the release of 
the kinetic energy frustration in the N\'eel phase. In this case, the kinetic frustration release is caused 
by the vanishing of the frustrating first neighbor hopping amplitudes in the $\bQ=(\pi,\pi)$ N\'eel state, and not by the appearance of a non-trivial 
spin-Berry phase like in the triangular lattice.

\subsection{Other lattices}

So far, we have found two examples where if the $t-J$ model with $J=0$ --equivalently, the $U=\infty$ Hubbard model-- is kinetically frustrated, 
there is a ground state degeneracy between a determined antiferromagnetic state and the ferromagnetic one at the mean field level. 
The selected antiferromagnetic state is the one which allows the release of the kinetic frustration. 
Our DMRG calculations allow us to affirm that in these cases, the antiferromagnetic state is the true ground state of the model, indicating that the 
phenomenon of kinetic antiferromagnetism seems to be quite ubiquitous in kinetically frustrated systems. Here, we briefly extend this kind of reasoning 
to other systems.

\begin{itemize}
\item 
We consider the mean-field analysis for the square lattice model with arbitrary $t_2/t_1$. 
We find that for any negative $t_2$ there is no kinetic frustration and, therefore, the Nagaoka ferromagnetic state is the unique ground state. 
On the other hand, for positive $t_2$ there is kinetic frustration. If $t_2 > |t_1|/2$ 
($t_1$ sign is irrelevant) the usual $(\pi,\pi)$ N\'eel and the ferromagnetic states are degenerate, while if $t_2 < |t_1|/2$ 
the ferromagnetic phase is the only ground state, with an energy difference relative to the N\'eel phase, $\Delta E \equiv E^{(\pi,\pi)}_{MF}-E^{\rm ferro}=2|t_1|-4t_2.$ 
Based in the previous examples, we expect that, in an exact calculation, the $(\pi,\pi)$ N\'eel phase will be the (only) ground state for $t_2 > |t_1|/2$. 
We can push further our prediction: the ferromagnetic energy is the exact one, while the antiferromagnetic energy can still decrease if we go beyond the 
mean-field approximation, and taking into account that previous DMRG computations give energy differences between the antiferromagnetic and the ferromagnetic states
of the order of $|t_1|$, we propose that the N\'eel state will be the ground state as long as $\Delta E \lesssim |t_1|$, that is, for $t_2 \gtrsim |t_1|/4$. 
Our preliminary DMRG results confirm this prediction, as we find that for $t_2/|t_1| \gtrsim 0.3$ the ground state has $(\pi,\pi)$ N\'eel order, while 
for $t_2/|t_1| \lesssim 0.3$ the ground state is ferromagnetic.

\item 
The anisotropic triangular lattice has hopping $t$ along the $\pm R_1=\pm (1,0)$ vectors, and $t'$ along the
$\pm R_2=\pm(\frac{1}{2},\frac{\sqrt{3}}{2})$ and $\pm R_3= \pm(-\frac{1}{2},\frac{\sqrt{3}}{2})$ vectors. For negative $t$, independently of the
$t'$ sign, the system is kinetically unfrustrated and the mean-field approach yields a ferromagnetic ground state, with an energy well separated from the
antiferromagnetic phases. For positive $t$, the kinetic energy is frustrated, and we find two different mean-field regimes: {\it i}) 
for $t > |t'|/2$ the ferromagnetic state is degenerate with a spiral phase characterized by a wave vector $\bQ=(Q,0),$ with $Q$ going continuously from 
0 ($|t'|=2t$) to $2\pi$ ($t'=0$); {\it ii}) for $t< |t'|/2$ the ferromagnetic phase is the only ground state. 
Drawing on arguments similar to those put forward in the $t_1-t_2$ square lattice model, we expect a kinetic antiferromagnetic 
ground state for $t \gtrsim |t'|/4$ and a ground state with ferromagnetic order for $t \lesssim \frac{|t'|}{4}$. 
Preliminary DMRG results confirm again this prediction, but the wave vector of the antiferromagnetic phases are different from the mean-field ones. 
DMRG yield a 120$^{\circ}$ N\'eel order for $t \gtrsim |t'|/2,$ while $\bQ=(Q,0)$ for $t \lesssim |t'|/2$ with $Q$ going from $Q=4\pi/3$ ($t \sim |t'|/2$) 
to $Q=0$ ($t \sim |t'|/4$). The prevalence of the 120$^{\circ}$ N\'eel phase would indicate that this order optimize the release of the kinetic energy frustration, 
but more work is needed to assert this affirmation.

\item 
Finally we consider the $U=\infty$ Hubbard model on the kagom\'e lattice. 
For negative $t$ there is no kinetic frustration, and the mean-field recovers the exact Nagaoka ground state, while
for positive $t$ we find that the ferromagnetic phase is degenerate with mean-field 
120$^{\circ}$ magnetic orders, like the $\sqrt{3}\times\sqrt{3}$ and ${\bq = 0}$ states \cite{harris92}. 
We speculate that, at the mean-field level, any state that obeys the condition of zero total spin at each triangle, that is, any classical ground state of the 
kagom\'e Heisenberg model, will be degenerate in energy with the ferromagnetic phase. We leave for future investigations the very interesting 
analysis of kinetic antiferromagnetism on the kagom\'e lattice. 
\end{itemize}


\begin{thebibliography}{99}
\bibitem{nota2} The expressions are $M_s = \sqrt{\frac{4S^{zz}(\bQ)}{N}}$  
and $M_s = \sqrt{\frac{S^{zz}(\bQ)}{N}}$ for the triangular and square lattices, respectively.
\bibitem{note3} To compare similar results between the mean field $M_s$ computed for the two dimensional 
triangular lattice, and DMRG computations done in clusters with $L_y=6$ legs,
we use the prescription given in Ref. \cite{white07}, about choosing the optimal cluster shapes. 
In the case of 6 legs, this cluster is the $9 \times 6$ ladder.
\end{thebibliography}
\begin{thebibliography}{99}
\bibitem{note1} In a finite cluster, the classical value is $B_R=\frac{1}{2}\left(1-\frac{1}{N}\right)\cos \frac{\bQ\bR}{2}.$ Here we are considering 
 the thermodynamic limit of this expression.
\end{thebibliography}
\end{document}